\documentclass[final,3p,times]{elsarticle}

\journal{High Energy Density Physics}

\begin{document}

\begin{frontmatter}

\title{Simulating the long-term evolution of radiative shocks in shock tubes}

\author[label1]{B. van der Holst\corref{cor1}}
\ead{bartvand@umich.edu}
\author[label1]{G. T\'oth}
\author[label1]{I.V. Sokolov}
\author[label2]{B.R. Torralva}
\author[label3]{K.G. Powell}
\author[label1]{R.P. Drake}

\address[label1]{Department of Atmospheric, Oceanic and Space Sciences,
  University of Michigan, Ann Arbor, MI 48109, USA}
\address[label2]{Materials Science and Engineering,
  University of Michigan, Ann Arbor, MI 48109, USA}
\address[label3]{Department of Aerospace Engineering,
  University of Michigan, Ann Arbor, MI 48109, USA}

\cortext[cor1]{Corresponding author}
\begin{abstract}

We present the latest improvements in the Center for Radiative Shock
Hydrodynamics (CRASH) code, a parallel block-adaptive-mesh Eulerian code for
simulating high-energy-density plasmas. The implementation can solve for
radiation models with either a gray or a multigroup method in the
flux-limited-diffusion approximation. The electrons and ions are allowed to be
out of temperature equilibrium and flux-limited electron thermal heat
conduction is included. We have recently implemented a CRASH laser package
with 3-D ray tracing, resulting in improved energy deposition evaluation. New,
more accurate opacity models are available which significantly improve
radiation transport in materials like xenon. In addition, the HYPRE
preconditioner has been added to improve the radiation implicit solver. With
this updated version of the CRASH code we study radiative shock
tube problems. In our set-up, a $1\,$ns, $3.8\,$kJ laser pulse irradiates a
$20\,$micron beryllium disk, driving a shock into a xenon-filled plastic tube.
The electrons emit radiation behind the shock. This radiation from the shocked
xenon preheats the unshocked xenon. Photons traveling ahead of the shock will
also interact with the plastic tube, heat it, and in turn this can drive
another shock off the wall into the xenon. We are now able to simulate the
long term evolution of radiative shocks.

\end{abstract}


\end{frontmatter}

\section{Introduction}

Radiative shocks are important in many astrophysical environments like
for instance supernova explosions, supernova remnants, and shocked molecular
clouds. With the emergence of high-energy-density (HED) facilities, such flows
can now be studied in detail in laboratory experiments. Laser-driven
experiments revealed many properties of radiative shocks.
Ionizing radiative precursors can appear ahead of strong shocks, see e.g.
Refs. \cite{keiter2002,bouquet2004,koenig2006}.
A radiative cooling layer was discussed for a system that is optically
thick in the shocked plasma, while optically thin in the unshocked material
\cite{reighard2007}. More recently,
magnetic fields were shown to be important in laser-produced shock waves.
Magnetic fields can for instance be generated by shock waves by
means of the Biermann battery process\cite{gregori2012}. Experiments are
proposed to produce collisionless shocks via counterstreaming plasmas
\cite{park2012}.
These laser-driven shock experiments do not only provide insight in the
underlying physical mechanism related to radiative shocks and the connections
to astrophysics. They are also useful for validating numerical simulation
codes that are used for the HED experiments and in the astrophysical context.

In recent radiative shock tube experiments \cite{doss2009,doss2010} it was
demonstrated that sufficiently fast shocks can produce wall shocks ahead
of the primary shock. These experiments were performed at the Omega
high-energy-density laser facility \cite{boehly1995} using ten laser beams
delivering a total energy of $3.8\,$kJ to a beryllium target. The ablated
beryllium drives like a piston a primary shock through a xenon-filled
plastic tube. This shock is sufficiently fast that it produces a radiative
precursor in the unshocked xenon, which will also heat the plastic tube
and subsequently can launch wall shocks. This compound shock is a challenging
problem for numerical simulation codes as it involves hydrodynamics and
radiation transport at different spatial and temporal scales. The center for
radiative shock hydrodynamics (CRASH) project aims to improve our
understanding of radiative shocks through experiments and simulations, and to
be able to predict radiative shock properties with a validated simulation code.

In previous modeling of radiative shock tubes with the CRASH code
\cite{vanderholst2011,vanderholst2012} we used the H2D simulation code, the
2D version of Hyades \cite{larsen1994} to evaluate the laser energy deposition
during the first $1\,$ns. H2D is Lagrangian radiation-hydrodynamics code with
multigroup radiation diffusion capability. After $1\,$ns we remapped the H2D
output to the CRASH code for further simulation. The latter code is
also a radiation-hydrodynamics code and uses the block adaptive tree library
(BATL) \cite{toth2012} to solve the equations on dynamically adaptive
Eulerian meshes. It currently includes flux-limited multigroup radiation
diffusion, flux-limited electron heat conduction, multi-material treatment with
equation-of-state and opacity solvers. While this suite of codes was
able to solve for the radiative shock structures in xenon-filled
nozzles \cite{vanderholst2012}, the more basic radiative shocks in xenon-filled
straight plastic tubes turned out to be problematic \cite{drake2011}.

In this paper, we describe several new improvements to the CRASH code. First
of all we have implemented a new parallel laser energy deposition library as an
integral part of our code. This allows the code to simulate the laser heating
and the subsequent radiation-hydrodynamic response in a self-consistent and
efficient way with
one single model. Another improvement was needed for the xenon opacities,
since the atomic data provided to our opacity solver were inaccurate.
We now use high quality xenon opacities calculated with the
super-transition-arrays (STA) model \cite{barshalom1989} as an alternative.
Both the introduction of the laser package and improved xenon opacities
turned out to make the radiation transport much stiffer in some regions.
The algebraic multigrid preconditioner using the BoomerAMG solver from
the HYPRE library \cite{falgout2002} resulted in more accurate solutions.
It is the purpose of this paper to demonstrate that these code changes
result in improvement in the fidelity of the simulation results.
The reported distortion of the compressed xenon layer on axis \cite{drake2011}
is now significiantly reduced and the wall shock is now more realistic.

The overarching goal of the CRASH project is to assess and improve the
predictive capability and uncertainty quantification of a simulation code,
using experimental data and statistical analysis \cite{holloway2011}. The
specific focus in this project is radiative shock hydrodynamics. The aim is
to predict the experimental radiative shock structure in elliptical nozzles
with our simulation code. To achieve this we first calibrate our code with
experimental results obtained for straight shock tubes and circular nozzles.
It is the purpose of this paper to demonstrate that we can now also model the
straight shock tube problem with sufficient fidelity that we can aim to
reproduce the experimental data.

The outline of this paper is as follows: Section \ref{sec:setup} describes
how we setup the shock tube and laser pulse with our new laser package that
is consistent with the experiments performed with the Omega laser facility.
This is followed in Section \ref{sec:results} by a discussion of the simulation
results. We conclude the paper in Section \ref{sec:conclusions}.

\section{Numerical setup of the shock tube experiment} \label{sec:setup}

In the baseline experiment of CRASH, a radiative shock is created by means
of ten laser beams from the Omega laser facility. The resulting laser pulse
irradiates a $20\,\mu$m beryllium target with approximately $3.8\,$kJ
laser light of $0.35\,\mu$m wavelength for the duration of $1\,$ns. This first
ablates the beryllium, generates a shock, and then accelerates the plasma to
over $100\,$km/s. The front of this plasma drives a shock through a
xenon-filled polyimide tube with an initial shock velocity of $200\,$km/s,
see also \cite{drake2011}. In the shocked xenon region, the shock-heated ions
exchange energy with the electrons so that they are also heated. The shock is
fast enough that the
energy balance requires a radiative cooling layer. The emitted photons from
this layer can propagate ahead of the shock and preheat the unshocked xenon.
A fraction of this radiation also expands sideways and heats the tube wall,
leading to ablation of the polyimide, which in turn drives a wall shock
into the xenon \cite{doss2009}. In this section, we will describe in more
detail how we numerically setup this experiment with our new laser package.

\begin{figure}
{\resizebox{0.48\textwidth}{!}{\includegraphics[clip=]{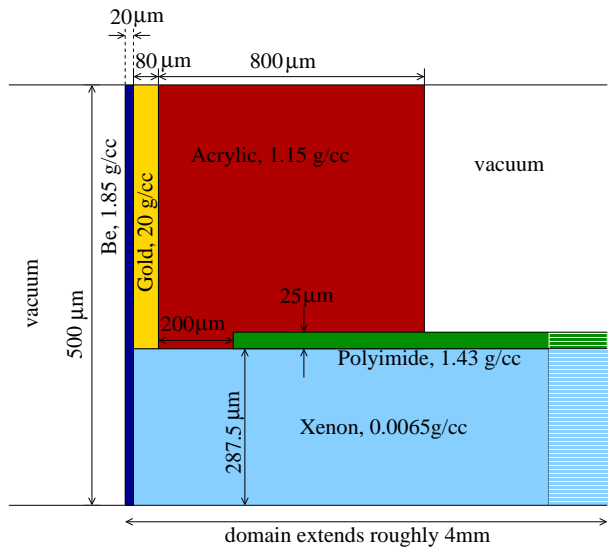}}}
\caption{Details of the axi-symmetric radiative shock tube target.}
\label{fig:crash_setup}
\end{figure}

The details of the base experiment is shown in the axi-symmetric plane
in Fig. \ref{fig:crash_setup}. The radial coordinate is in the vertical
direction, while the tube direction is horizontally and the axis of symmetry
is at the bottom. The $20\,\mu$m thick beryllium disk on the left is the
target and the laser light will come in from the left of this disk. The
beryllium disk is attached to a $4\,$mm long straight polyimide tube with
$287.5\,\mu$m inner radius and $25\,\mu$m wall thickness. This tube is filled
with xenon gas with an initial mass density of $0.0065\,$g/cm$^3$ and a
pressure of $\sim 1.2\,$atm. The tube is used to prevent the radiative shock
from expanding and hence to prevent the slowing down of the shock such that it
is no longer radiative. The interesting side effect of this tube is the
aforementioned wall shock. A gold washer is placed directly behind the
beryllium disk to prevent laser-driven shocks outside the polyimide tube.
We use acrylic in between the gold washer and the polyimide tube. The vacuum
outside the tube target is replaced in the simulation by very low density
polyimide to prevent
zero mass density in the numerical schemes. The vacuum to the left of the
beryllium disk is replaced by beryllium with a density much smaller than the
critical density such that this beryllium does not impact the laser energy
deposition.

We currently use levelset functions to track the different materials in time.
To initialize these levelset functions, the interfaces between the materials
are defined as a number of segment vectors with the materials on the left and
the right, see \ref{sec:interface} for some algorithmc details. Based on these
material interfaces we construct the levelset
functions as smooth and signed distance from the material interface, in which
the sign is positive inside the material region, while negative outside.
With these levelset functions we can then reconstruct the type of material
in each computational cell.

The simulations in this paper have been performed with the new 3-D laser
package described in \ref{sec:laser}. The numerical set-up is as
follows: A laser pulse of $0.35\,\mu$m wavelength irradiates a
$20\,\mu$m thick beryllium disk for $1\,$ns. The corresponding critical
electron number density, below which all light is absorbed, is
$8.9\times10^{21}\,$cm$^{-3}$. In the radiative shock experiments, the total
laser energy deposition is typically $3.8\,$kJ, but in our
simulations we have to scale this down to arrive at similar results in the
shock position. This laser scale factor accounts for that part of the
energy for which the laser-plasma interactions cause reflection or absorption
into distributions of particles that do not effectively generate ablation.
These laser-plasma processes include wave-wave instabilities and related
phenomena. \cite{kruer2001} In this paper, we will use an energy of $2.7\,$kJ.
The laser spot size is $820\,\mu$m full width half maximum (FWHM) diameter.
In our application, the Omega laser pulse can be represented
by 10 beams with a circular cross-section and the following angles with
respect to the shock tube axis:
$10.13\,^{\circ}, 10.49\,^{\circ}, 31.37\,^{\circ}, 31.6\,^{\circ},
31.71\,^{\circ}, 31.94\,^{\circ}, 41.96\,^{\circ}, 42.04\,^{\circ}, 
42.37\,^{\circ}, 50.62\,^{\circ}$. For computational efficiency, this is
modeled using 4 beams with the power per beam weighted by the number
of beams approximately at that angle: one beam at $50.6\,^{\circ}$,
three beams at $42.0\,^{\circ}$, four beams at $31.7\,^{\circ}$, and
two beams at $10.2\,^{\circ}$. The laser profiles are spatially chosen as
super-Gaussian of the order $4.2$ and the time profile is split in a $100\,$ps
linear ramp-up phase, $0.8\,$ns with constant power, and a $100\,$ps linear
decay time. Each beam is discretized with 900 by 4 rays for the radial and
angular coordinates of the beam cross-section. The radial beam domain size is
up to 1.5 times the FWHM beam radius of $410\,\mu$m, while the angular
direction is
limited to half the domain $\left[ 0, \pi \right]$ due to symmetry
considerations. The resulting beam resolution is sufficiently high to obtain a
smooth laser heating
profile, but also as coarse as possible for computational speed.

\section{Radiative shocks in straight tubes}\label{sec:results}

We use the CRASH code \cite{vanderholst2011} to simulate both the laser energy
deposition and the radiative shock propagation. This code solves the
multi-material radiation-hydrodynamic equations in an operator split fashion.
For each time step, we split the dynamical equations in the following way:
(1) The hydrodynamic equations, level sets, and advection of radiation groups
are explicitly solved with a shock capturing scheme. We typically use the
HLLE scheme with a Courant--Friedrichs--Lewy (CFL) number of 0.8 and the
generalized Koren limiter with $\beta=3/2$. (2) Optionally we add a
frequency advection in the radiation group energies caused by fluid
compression. In this paper, this is switched off. (3) The contribution of the
laser heating is explicitly added
to the electron internal energy. In our code this energy is split between
electron pressure and an extra internal energy to account
for EOS corrections like the ionization, excitation, and Coulomb interactions
of partially ionized ion-electron plasma. (4) The radiation diffusion, heat
conduction, and energy exchanges are solved implicitly. We use a multigroup
flux-limited radiation diffusion method in which the flux limiter is the
square-root flux limiter \cite{morel2000}. For the electron thermal
heat conduction, we use the so-called threshold model \cite{drake2011}, for
which the heat flux limiter has a value of $0.06$. We have made several
improvements to the implicit solver including the implementation of the HYPRE
preconditioner, as described in \ref{sec:hypre}, to make the solutions more
accurate. The results demonstrated below have been produced with these new
code changes.

The photon energy range in our multigroup radiation model is
$0.1\,$eV -- $20\,$keV, which is divided in 30 groups.
These groups are non-logarithmically distributed to improve the accuracy for
absorption edges in the used materials. In our code, the frequency-dependent
absorption coefficients are calculated internally and include the effects of
Bremsstrahlung, photo-ionization of the outermost electrons, and bound-bound
transitions with spectral line broadening. Multigroup opacities are then
determined by averaging the absorption coefficients over the photon energy
groups. The resulting specific Rosseland and Planck mean opacities for all
groups are stored in lookup tables. For xenon opacities, the available atomic
data provides a description of the actual structure that is too incomplete that
the methods used by our code produce substantially inaccurate opacities. We
therefore use for xenon high quality opacity tables calculated with the
super-transition-arrays (STA) model \cite{barshalom1989} of Artep, Inc.
In the relevant ranges of xenon densities and temperatures, STA produces
higher opacity values than we find by running the CRASH opacity model using
the limited available atomic data, e.g. for xenon at a temperature of
$49.99\,$eV and a density $\rho = 0.011\,$g/cm$^3$ the STA opacities
around $100\,$eV photon energies are three orders of magnitude larger.
The STA Fe and Ni opacities have recently been compared with other models in
Ref. \cite{gilles2011}.

The shock tube is defined on a 2-D axial symmetric computational domain. The
size of the domain is $-550 < x < 3850$ along the tube and the radius is
limited to $0<r<500$, with all distances measured in microns. The base level
grid is decomposed of $165\times15$ grid blocks of $8\times8$ mesh cells for
the $x$ and $r$ directions, respectively. Two levels of adaptive mesh
refinement
are applied, so that the effective grid resolution is $5280\times480$ grid
cells of approximately $0.8\,\mu$m by $1\,\mu$m. The grid refinement is
applied at all interfaces involving xenon or gold. We also apply grid
refinement when the xenon mass density exceeds $0.02\,$g/cm$^3$ in order to
resolve the xenon shock front, the electron-ion equilibration zone, and the
radiative cooling layer in the shocked xenon. In addition, all
beryllium to the right of $x=-5\,\mu$m is mesh refined during the laser
heating. All mentioned grid refinements are applied when any of the
mentioned criteria is met in the mesh as well as ghost cells. Note that
the effective cell size of $0.8\,\mu$m means that there are 25
cells in the $x$-direction to span the beryllium disk thickness. This turns
out to be sufficient to accurately describe the beryllium ablation and
shock breakout.

The boundary conditions at the symmetry axis $r=0$ are reflective, while we
use at all other boundaries extrapolation with zero gradient. However, for the
radiation groups, we use a zero incoming flux boundary condition at the
outer boundaries, i.e. all radiation leaving the computational domain will not
return back.

The simulated laser energy deposition and radiative shock evolution is modeled
from $0$ to $18\,$ns. During the first $200\,$ps the time step is reduced
from about $3\times10^{-17}\,$s at the very beginning of the simulation and
gradually increases towards the end of the $200\,$ps to a time step based on a
CFL of 0.8. The increase (or decrease) in time step is controlled by the
change in the extra internal energy that accounts for the ionization,
excitation, and Coulomb interactions. From $200\,$ps to $18\,$ns, the
time step is set by the default CFL number.
This computation was performed on 100 processors of the
FLUX supercomputer at the University of Michigan using dual socket six-core
Intel Core I7 CPU nodes connected with infiniband and took $39.5\,$hours.
This includes $10\,$hours and $50\,$minutes for the laser heating,
$21\,$ hours for the Krylov solver, and $6\,$hours for the
setup time of the HYPRE BoomerAMG preconditioner.

\begin{figure}
{\resizebox{0.96\textwidth}{!}{\includegraphics[clip=]{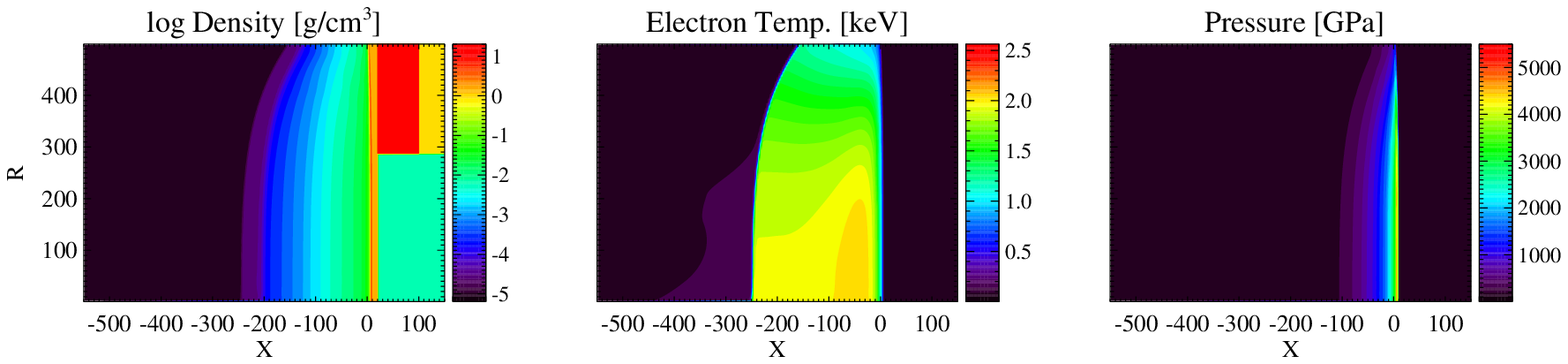}}}
{\resizebox{0.96\textwidth}{!}{\includegraphics[clip=]{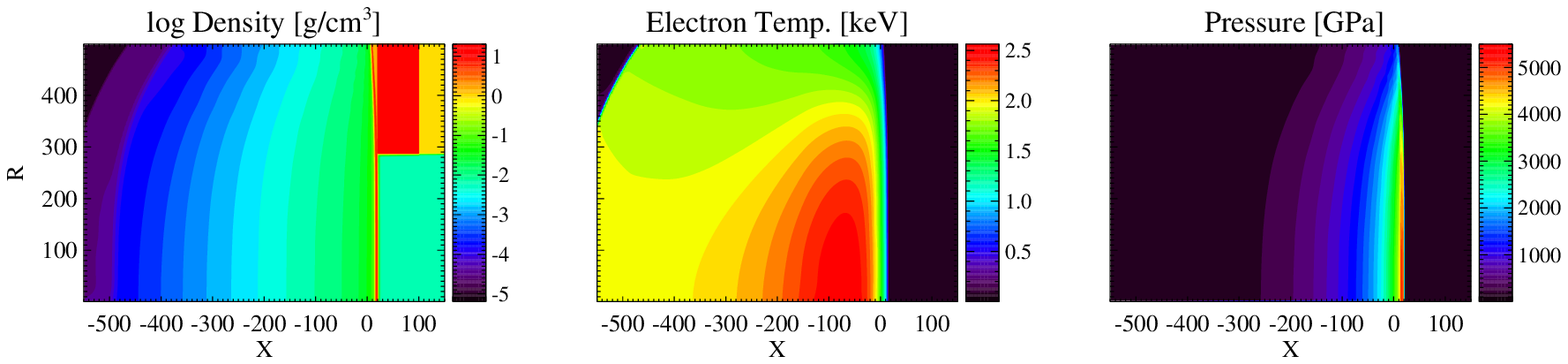}}}
\caption{The mass density (left panels), electron temperature (middle panels)
and total plasma pressure (right panels) as a function of the $x$ and $r$
coordinates in microns at time $t=200\,$ps (top row) and near the shock
breakout time $t=400\,$ps (bottom row).}
\label{fig:earlytime}
\end{figure}

In Fig. \ref{fig:earlytime} we show the early time response to the
laser heating. The top row is for the density, electron temperature and
total plasma pressure at time $t=0.2\,$ns. Here one can see the early
ablation of the beryllium disk, initially located between $x=0$ and
$x=20\,\mu$m. The bottom row shows the same but at time $t=0.4\,$ns.
The region to the left of the beryllium disk is the laser corona. The region
can be split in three main regions \cite{drake2006}: (1) The leftmost low
density region is the so-called expansion region in which the plasma expands,
but hardly absorbs the laser light. (2) Between the expansion region and the
critical density is the absorption region where the laser energy deposition
takes place. (3) Between the critical density and the not yet ablated
beryllium disk is the transport region in which electron heat conduction
transports heat from the low density and hot laser corona to the high density
and low temperature beryllium target. It is in this region that the
classical Spitzer-H\"arm (SH) formalism for heat transport overestimates the
heat flux for the steep temperature gradients. The artificial heat
flux limiter is used to prevent the SH heat conduction from exceeding the
the free-streaming heat flux (see also \cite{drake2006}).

The resulting ablation pressure of approximately $5000\,$GPa drives a shock
through the beryllium disk. At $t \approx 400\,$ps the shock has reached the
right boundary of the beryllium disk. This is the shock breakout
time. After this time the high pressure due to the laser heating will further
accelerate the shocked plasma. In a forthcoming paper, we will present a
more detailed analysis and compare the simulated shock breakout with
experiments.

\begin{figure}
\begin{center}
{\resizebox{0.48\textwidth}{!}{\includegraphics[clip=]{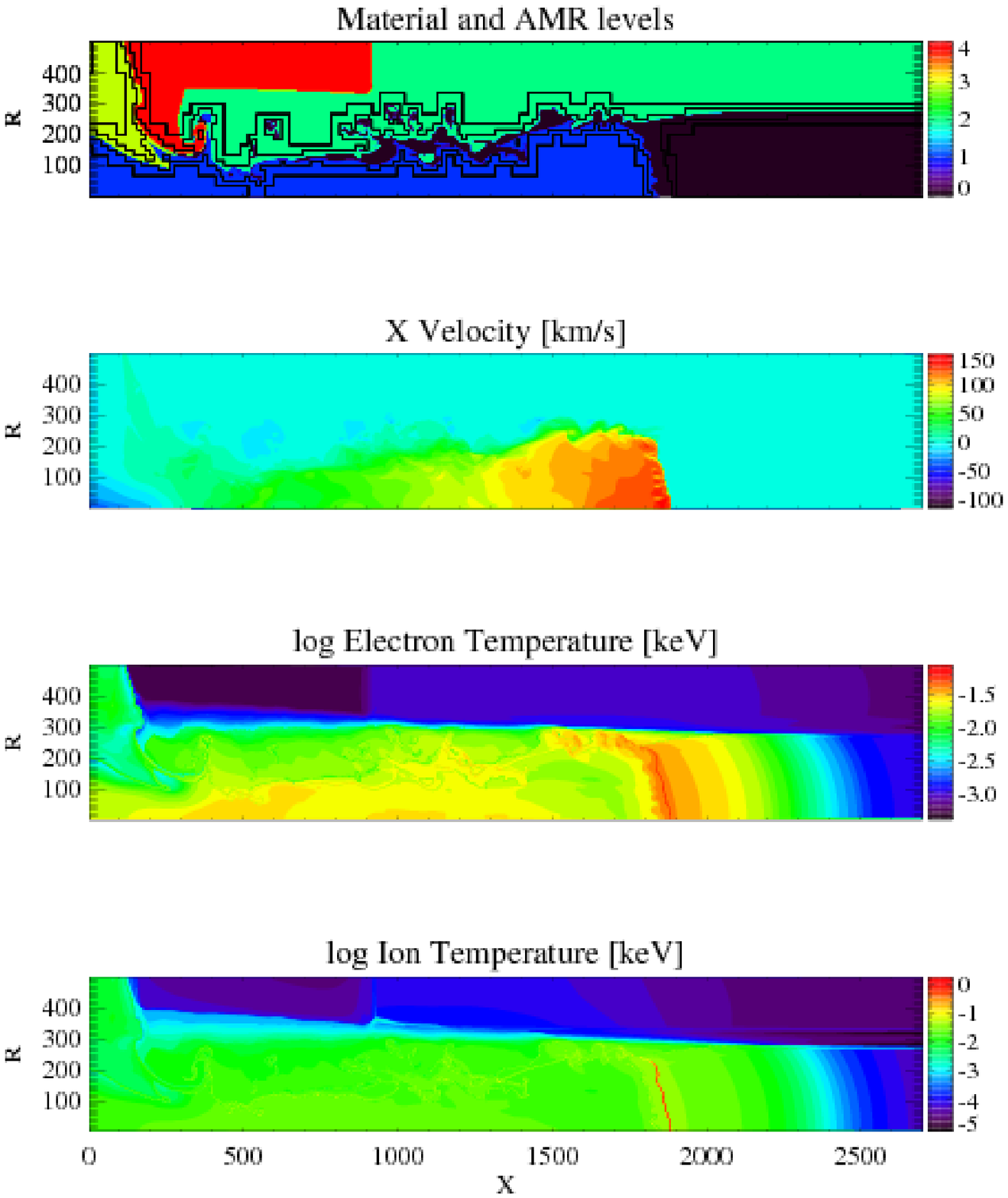}}}
{\resizebox{0.48\textwidth}{!}{\includegraphics[clip=]{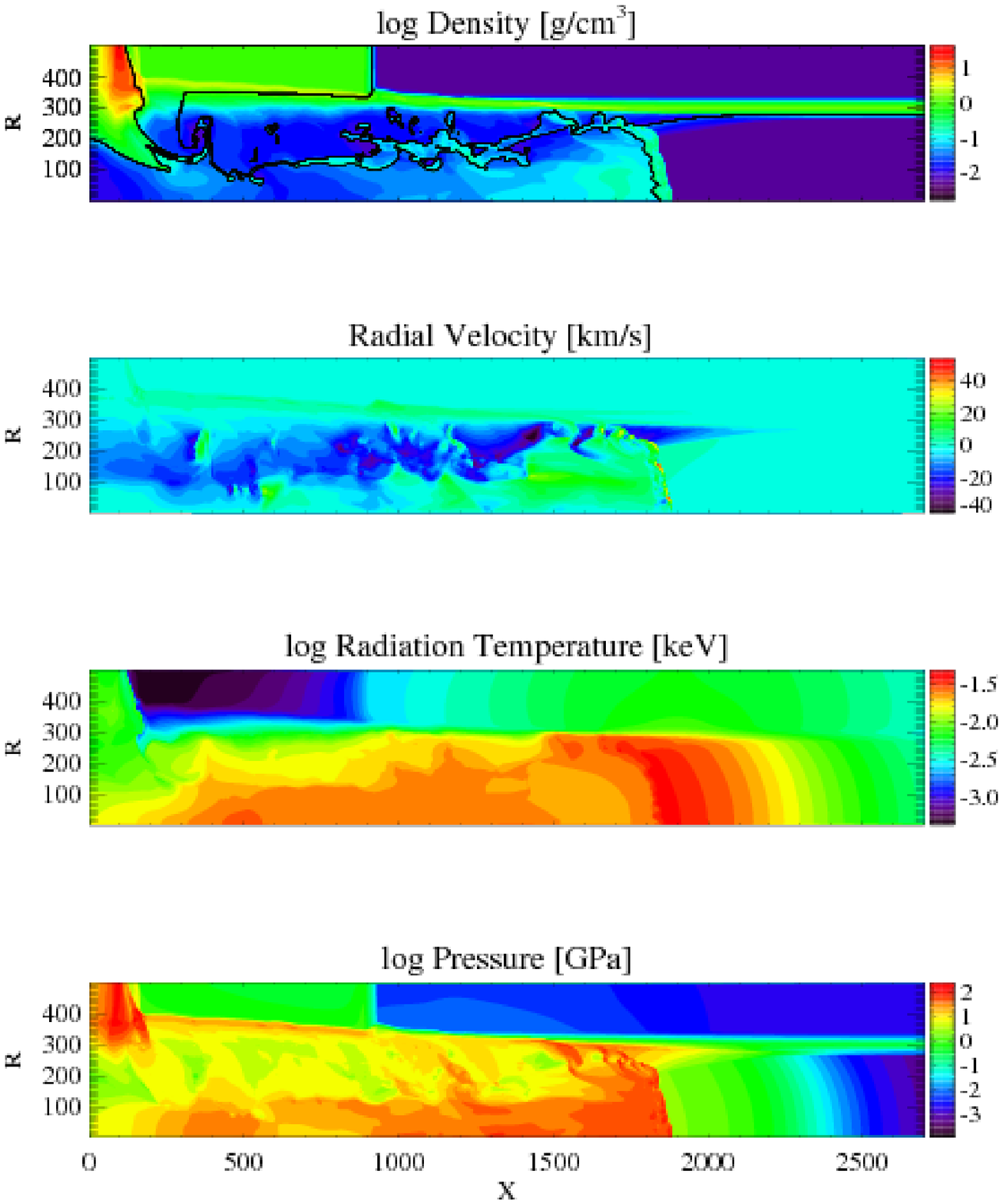}}}
\end{center}
\caption{The radiative shock structure at $13\,$ns. The colors in top-left
panel indicate the materials: xenon
(black), beryllium (blue), gold (yellow), acrylic (red) and polyimide (green),
while the black lines indicate grid resolution changes.}
\label{fig:state_13ns}
\end{figure}

The shock structure at $13\,$ns is shown in Fig. \ref{fig:state_13ns}. In the
top left panel the materials are displayed: xenon (black), beryllium (blue),
gold (yellow), acrylic (red), polyimide (green). The two levels of dynamic mesh
refinement are indicated by the black lines. The beryllium moves through the
polyimide tube and drives like a piston a shock into the xenon. The compressed
xenon between this shock front and the beryllium is found around
$x=1850\,\mu$m in the mass density plot of the top right panel. The plot
also shows, via a black line, where the material interfaces are. The
physics described in the remaining panels is similar to that in previous
studies, see Ref. \cite{vanderholst2012}. We repeat here only the main
results for completeness. The shock velocity at $13\,$ns has gradually reduced
from the early velocity of $200\,$km/s to about $150\,$km/s. The
bottom right panel shows the pressure jump at the shock, while the bottom
left panel shows that the ions are shock heated. In the compressed xenon
region this leads first to an electron-ion temperature equilibration
due to Coulomb collisions, resulting in a cooling of the ions and heating of
the electrons. Further to the left in the compressed xenon region, the
electrons cool down by emitting photons. This is called the radiative
cooling layer \cite{reighard2007}. The emitted photons can propagate ahead
of the shock and produce a radiative precursor as depicted in the
radiation temperature panel. The sideways propagation of the radiation heats
the polyimide tube. The ablation pressure of the polyimide at
$x\approx 2000\,\mu$m in the pressure plot then drives a wall shock radially
inward as is visible at the same $x$ location in the density and radial
velocity plots.

The main goal of the present paper is to demonstrate that with the new improved
physics fidelity and numerical schemes since the CRASH code release of
Ref. \cite{vanderholst2011}, the radiative shock simulations
in straight tubes are no longer susceptible to distortion of the dense xenon
layer on axis. Indeed, the primary shock front Fig. \ref{fig:state_13ns}
is nearly straight and only slightly slanted, but does not display the unwanted
protrusion of the shock as shown in \cite{drake2011}. The code change that
is instrumental in the improved radiative shock front is the new laser
package instead of using the H2D code for the initial 1.1ns.

\begin{figure}
\begin{center}
{\resizebox{0.48\textwidth}{!}{\includegraphics[clip=]{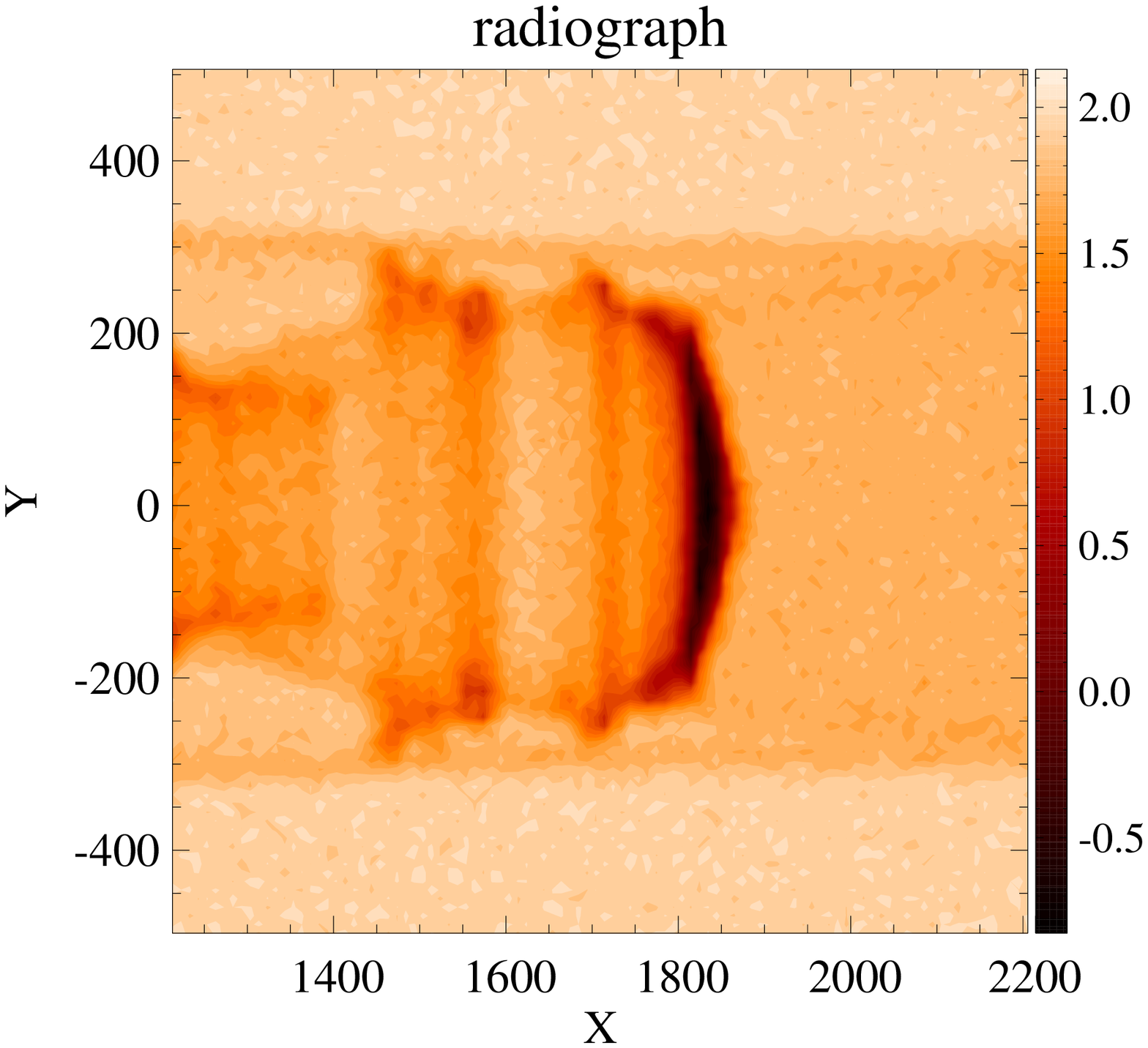}}}
{\resizebox{0.4\textwidth}{!}{\includegraphics[clip=]{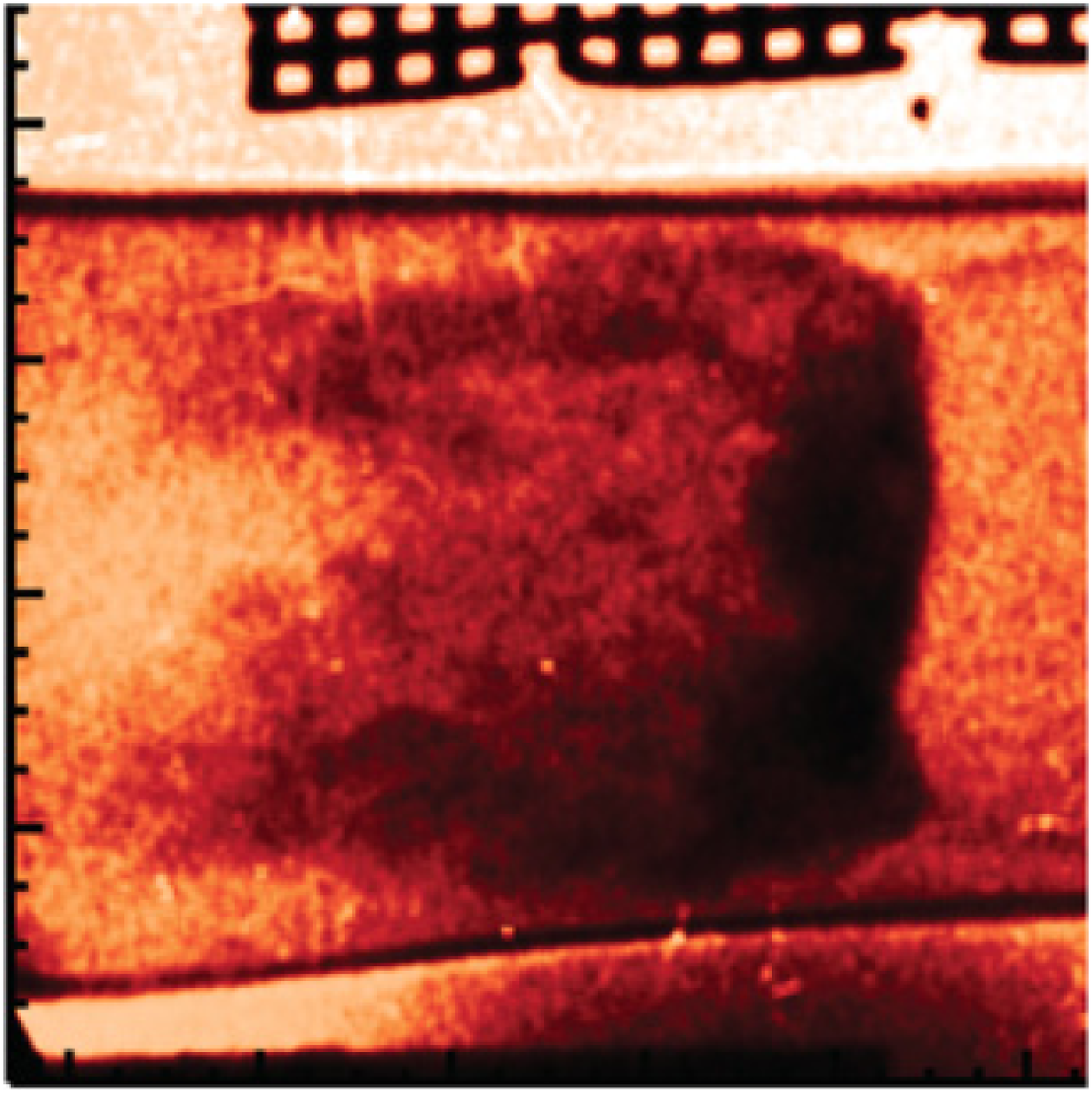}}}
\end{center}
\caption{(left panel) Simulated radiograph image at $13\,$ns. (Right panel)
Experimental x-ray radiograph from \cite{doss2010}. The experimental and
simulated set-up is not identical and hence shock positions are different.}
\label{fig:radiograph}
\end{figure}

From the radiative shock tube experiments we obtain backlit-pinhole
radiograph images \cite{doss2009}. These images are produced by transmitting
$5.18\,$keV through the CRASH target and in essence show regions of dense
xenon. From these image we can then deduce the location of the primary and
wall shock. With our code we can produce simulated X-ray radiographs, see the
left panel of Fig.
\ref{fig:radiograph}, and use such images in future code validation and
uncertainty quantification. The importance of the code improvements, reported
in the present paper, is that there is no longer dense xenon in front
of the center of the primary shock at $(x,y)\approx(1850,0)\,\mu$m and hence
there is no longer a dark feature ahead of the dense xenon layer in the
radiograph as in the experimentally obtained radiograph in the right
panel of Fig. \ref{fig:radiograph}.

\begin{figure}
{\resizebox{0.48\textwidth}{!}{\includegraphics[clip=]{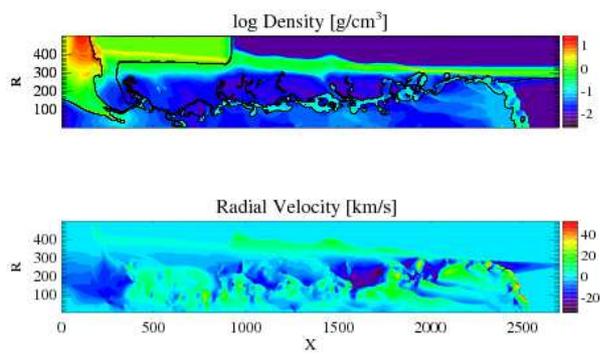}}}
\caption{The shock structure at 18 ns.}
\label{fig:state_18ns}
\end{figure}

To demonstrate that the shock is also correct at later times, the density
and radial velocity at time $18\,$ns is shown in Fig. \ref{fig:state_18ns}.
The primary shock has reached $x\approx 2500\,\mu$m and is somewhat more
slanted. The radially inward moving wall shock is at the far right in these
panels.

\section{Conclusions}\label{sec:conclusions}

In this paper we discussed laser-driven radiative shocks in xenon-filled
straight plastic tubes. The simulations capture the shock properties
seen in the laser-driven experiments and radiation-hydrodynamic theories.
The laser heating ablates the beryllium target, which subsequently drives
a shock through a xenon-filled tube. For laser energies of $3.8\,$kJ,
scaled down to $2.7\,$kJ,
beryllium disk thickness of $20\,\mu$m and initial xenon gas pressure of
$1.2\,$atm we obtain initial shock velocities of $200\,$km/s. These shocks
are fast enough to produce a radiative cooling layer in the shocked xenon.
The photons can propagate ahead of the primary shock and produce a
radiatively heated precursor in the unshocked xenon. The radiation also
expands sideways and heats the plastic wall ahead of the primary shock.
This produces an inward moving wall shock.

The numerical modeling was performed with the CRASH simulation code that
solves the radiation-hydrodynamic equations on a block adaptive Eulerian grid.
The hydrodynamic equations are solved explicitly, while the radiation
diffusion, electron heat conduction and energy exchanges are treated
implicitly. This code also includes equation-of-state and opacity solvers to
enable multi-material simulations. Several code improvements have been
implemented that enabled the improved quality of straight shock tube
simulations. A new laser package with 3-D ray tracing has been added to CRASH
so that the simulations can be performed self-consistently in one single model
instead initializing the simulation runs with an external lagrangian code
to evaluate the laser-energy deposition.  We also incorporated highly
accurate xenon opacities calculated with the super-transition-arrays (STA)
model. Finally, we improved
the robustness of the implicit radiation solvers with the HYPRE preconditioner
instead of the original Block Incomplete Lower-Upper decomposition
preconditioner. These developments greatly improved the fidelity of the
radiative shock tube results.

From both the experimental and simulated X-ray radiographs, we can deduce
quantities like primary shock position, wall shock position, and compressed
xenon layer thickness. In future work, we will employ these type of metrics
for code validation. The goal of our project is to use the validated CRASH
code to perform predictive studies of three-dimemsional radiation-hydrodynamic
flows, such as radiative shocks in elliptical nozzles.

\section*{Acknowledgments}

This work was funded by the Predictive Sciences Academic Alliances Program in
DOE/NNSA-ASC via grant DEFC52-08NA28616 and by the University of Michigan.

\appendix

\section{Laser energy deposition library}\label{sec:laser}

In this appendix, we present a new package that models the laser energy
transport and deposition. In previous reported
radiative shock tube modeling efforts \cite{vanderholst2011,vanderholst2012},
we used the Lagrangian radiation-hydrodynamics code H2D \cite{larsen1994} for
the first $\sim1.1\,$ns. This initializing of the CRASH simulations with H2D
turned out to be problematic in that this produced significantly different
shock structures when compared to the observations \cite{drake2011}. We
therefore opted to implement our own laser package directly into the CRASH
code. This also allows us now to simulate the radiative shock experiment in a
single self-consistent model. In addition, the new laser package is parallel,
while H2D is a serial code, resulting in improved computational speed.

This laser package model decomposes the laser pulse into many rays. We use
a ray-tracing algorithm based on a geometric optics approximation with
laser absorption via inverse Bremsstrahlung along the trajectory of the rays.
Geometric optics is acceptable as long as the electron density does not vary
significantly over one wavelength of the laser pulse. This is satisfied
most of the time for our applications of interest, with the exception of
the startup phase of the laser heating. The inverse Bremsstrahlung absorption
is the most important absorption mechanism for the CRASH laser applications
\cite{drake2006}.

The laser package works both in the 2-D axi-symmetric geometry as well as
in 3-D cartesian. For the axi-symmetric geometry we have implemented two
versions of the the ray tracing: (1) rays confined to the axi-symmetric plane
and (2) ray tracing in 3-D. In the 2-D ray tracing case we experienced the
problem that all rays that are not parallel to the cylindrical axis will
eventually also heat the plasma near the cylindrical axis, resulting in an
excessive increase of the electron temperature near the axis. Using 3-D ray
tracing in the axi-symmetric geometry mitigates this problem, resulting in
improved simulations of shock break-out time and evolution of the radiative
shocks compared to the 2-D rays.

Our computationally parallel ray-tracing algorithm is based on previous
work on tracing radio rays in the solar corona \cite{benkevitch2010}.
Here, we briefly summarize the implementation as needed for the laser
heating. At each time step and for each ray we trace the trajectory with a
ray equation that can be derived from Fermat's principle: a ray
connecting two points ${\bf r}_1$ and ${\bf r}_2$ will follow a path
which minimizes the integral of refractive index $n$, i.e.
the variation of the integral
\begin{equation}
  \delta\int_{{\bf r}_1}^{{\bf r}_2} n({\bf r}) ds = 0,
\end{equation}
where the independent variable $s$ is the arc-length of the ray. This can be
shown, see Ref. \cite{benkevitch2010}, to be equivalent to
\begin{equation}
  \frac{d}{ds}\left( n\frac{d{\bf r}}{ds}\right) -\nabla n = 0,
  \quad {\rm or}, \quad
  \frac{d{\bf r}^2}{ds^2} = \frac{d{\bf r}}{ds}\times\left(
  \frac{\nabla n}{n}\times\frac{d{\bf r}}{ds}\right). \label{eq:ray}
\end{equation}
By introducing the ray direction, ${\bf v} = d{\bf r}/ds$, the system
(\ref{eq:ray}) of three second-order differential equations is transformed in a
set of six first-order equations
\begin{eqnarray}
  \frac{d{\bf r}}{ds} &=& {\bf v}, \label{eq:ray1}\\
  \frac{d{\bf v}}{ds} &=& {\bf v}\times\left(
  \frac{\nabla n}{n}\times {\bf v} \right) \label{eq:ray2}.
\end{eqnarray}
For isotropic collisionless plasmas the refractive index is
\begin{equation}
  n^2 = \varepsilon = 1 - \frac{\omega_p^2}{\omega^2},
\end{equation}
where $\varepsilon$ is the dielectric permittivity of the plasma,
$\omega$ is the frequency of the laser light, and the plasma frequency
$\omega_p = \sqrt{e^2n_e/m_e\varepsilon_0}$ depends on the electron density
$n_e$, electron mass $m_e$, electron charge $e$ and the permittivity of
vacuum $\varepsilon_0$. The refraction index is therefore determined from
the mass density via $n^2 = 1 - Z\rho/\rho_{\rm c}$, in which $Z$ is the ion
charge and the critical mass density is defined as
\begin{equation}
  \rho_{\rm c} = \frac{\varepsilon_0 A m_p m_e \omega^2}{e^2},
\end{equation}
and where $A$ is the mean atomic weight and $m_p$ is the proton mass. This
provides us with a final expression for the relative gradient of the
refractive index
\begin{equation}
  \frac{\nabla n}{n} = - \frac{\nabla(\rho Z)}{2(\rho_{\rm c} - \rho Z)}.
\end{equation}
Once this gradient is known, the integration of Eqs.
(\ref{eq:ray1})--(\ref{eq:ray2}) is performed with CYLRAD algorithm
\cite{boris1970}. The ray trace algorithm of
\cite{benkevitch2010} is implemented with adaptive step size to handle the
steep gradients in the plasma density. For each integration step, every ray is
checked for accuracy and correctness, and we ensure that rays do not
penetrate in regions where $Z\rho>\rho_c$.

Electron-ion collisions modify the refractive index to a complex value, where
the imaginary part corresponds to absorption
\begin{equation}
  n^2 = \varepsilon = 1 - \frac{\omega_p^2}{\omega(\omega + i\nu_{\rm eff})}.
\end{equation}
The effective electron-ion collision frequency is defined as
\cite{ginzburg1964}
\begin{equation}
  \nu_{\rm eff} = \frac{4\pi}{3} \sqrt{\frac{2k_BT_e}{\pi m_e}}
  \left( \frac{e^2}{k_BT_e}\right)^2 \left<n_i Z^2\right> \ln\Lambda.
\end{equation}
Here $n_i$ is the ion number density, $k_B$ is the Boltzmann constant,
$T_e$ is the electron temperature, and $\ln\Lambda$ is the Coulomb
logarithm. It is due to these electron-ion collisions that the laser energy
is absorbed into the plasma. The absorption coefficient (in units of 1/m) is
then found as \cite{drake2006}
\begin{equation}
  \alpha = \frac{\nu_{\rm eff}}{c} \frac{Z\rho/\rho_c}{1-Z\rho/\rho_c}.
\end{equation}
While performing the integration along each ray, energy is gradually deposited
in the plasma.

\begin{figure}
{\resizebox{0.48\textwidth}{!}{\includegraphics[clip=]{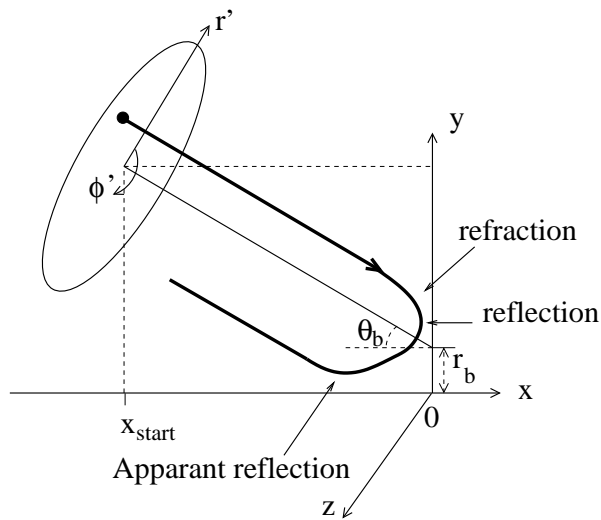}}}
\caption{Each laser beam has a different direction $\theta_b$ relative to the
$x$-axis. The beam rays start at $x_{\rm start}$ and $r_b$ defines the initial
$y$-position. The beam irradiance profile and the initial beam ray locations
are defined as a function of the polar beam coordinates $r'$ and
$\phi'$. The thick drawn line is a 3-D ray projected onto the axi-symmetric
plane, here shown as the $xy$-plane.}
\label{fig:laserbeam}
\end{figure}

We have added code infrastructure in CRASH to facilitate the setup of a laser
pulse using a set of rays. For the sake of brevity we will only describe the
3-D laser pulse implementation in a 2-D axi-symmetric simulation.
In general a 3-D laser pule will generate a 3-D laser heating, and
hence 3-D simulations are required. We have code infrastructure to perform
such 3-D simulations, but this is computationally expensive.
In the present paper we assume that the departure
from axi-symmetry is small. The laser pulse is defined with an
irradiance (in units of $J/s$) and a time profile with a given linear ramp-up
time, decay time, and total pulse duration. The laser pulse is further
decomposed in a number of laser beams with a circular cross-section. Each beam
is defined by a slope $\theta_b$ with respect to the $x$-axis as in Fig.
\ref{fig:laserbeam} and an initial starting point of the beam defined by
$x_{\rm start}$ and a beam offset $r_b$ from the $x$-axis. For the beams we
select a spatial profile of irradiance that is super-Gaussian as a function of
the beam radius $r'$: $\propto\exp(-(r'/a)^b)$,
where $a$ is the beam radius and $b$ is the super-Gaussian order usually
chosen to be $4.2$. We use a beam margin of $r'=1.5a$, beyond which the laser
energy deposition is assumed to be negligible. Each beam is discretized
with the same number of rays $n_{r'}$ for the radial beam direction and
$n_{\phi'}$ rays for the angular direction $\phi'$. Due to symmetry properties
for 3-D beams in axi-symmetric geometry, we only need to consider half of the
angular direction of the beam, i.e. we limit to the angular beam range
$\left[ 0,\pi \right]$. The number of radial rays should be high enough to
obtain a smooth laser heating profile, but as low as possible for computational
speed. This number $n_{r'}$ is typically chosen such that every computational
cell at the critical density surface, $Z\rho=\rho_c$, is crossed by at least
one ray, preferably more. If the
slopes $\theta_b$ of the laser beams are not too large, the number of angular
rays $n_{\phi'}$ can be selected to be small for the sake of computational
speed, while still achieving sufficient accuracy.

For the laser energy deposition, we evaluate the energy emitted by the laser
at each time step. The beam energy is distributed over the number of rays. The
local intensity for each ray is obtained from the super-gaussian beam
profile. The laser energy transport in the CRASH code uses 3-D
ray-tracing based on geometric optics. During the 3-D ray tracing, we use
the $x$ and $r=\sqrt{y^2+z^2}$ positions in the axi-symmetric plane to
determine where in the solution plane the laser energy deposition via inverse
bremsstrahlung occurs as well as the further evaluation of the trajectory in
3-D. For these evaluations we need to map the density and density gradient
into the 3-D space. A drawing of a 3-D ray projected onto the axi-symmetric
plane is shown in Fig. \ref{fig:laserbeam}. The projected ray shows not only
the reflection before approaching the critical density surface and the
refraction, but also the apparant reflection at a finite distance from the
$x$-axis instead of a reflection on axis. This apparant reflection is
because the 3-D ray is in general not in the axi-symmetric plane and hence
has a minimum distance with respect to the $x$-axis. It is this effect that
avoids the excessive laser heating on axis as is the case for 2-D rays in
which the rays are confined to the axi-symmetric plane.
The deposited laser energy at the current ray location in the
axi-symmetric plane is distributed to
the nearest computational zone volumes with the sum of the interpolation
coefficents equal to one and subsequently added as an explicit source term to
the right-hand-side of the electron energy density equation. The scheme is
fully conservative since the total energy that is deposited by the laser
pulse equals the total laser energy that is absorbed by the plasma.

\section{Improved radiation solver with HYPRE}\label{sec:hypre}

In this appendix we describe the improvements in the radiation diffusion
and heat conduction solvers of CRASH which led to a more robust numerical
scheme. These changes were especially needed due to introduction of the
laser package and the more realistic STA xenon opacities, which made the
thermal heat and radiation transport stiffer in some regions.

In the applications relevant to CRASH, the electron temperature does not
change much in the energy exchange between the electrons and radiation. We can
therefore first solve for the electron and ion temperature, $T_e$ and $T_i$,
without taking into account the energy exchange with the radiation.
Discretizing the heat conduction and electron-ion energy exchange in time
leads to the following backward Euler scheme:
\begin{eqnarray}
  C_{Vi}^*\frac{T_i^{n+1}-T_i^*}{\Delta t} &=&
  \sigma_{ie}^*(T_e^{**} - T_i^{n+1}), \label{eq:splitionenergy}\\
  C_{Ve}^*\frac{T_e^{**}-T_e^*}{\Delta t} &=&
  \sigma_{ie}^*(T_i^{n+1} - T_e^{**}) + \nabla\cdot C_e^*\nabla T_e^{**},
  \label{eq:splitelectronenergy}
\end{eqnarray}
where $C_{Vi}$ and $C_{Ve}$ are ion and electron specific heat, $C_e$
is the heat conduction coefficient, and $\sigma_{ie}$ is the coupling
coefficient that depends on the ion-electron relaxation time. These
coefficients are frozen in at time level $*$ during the time advance
$\Delta t$ from time level $*$ to $n+1$. The scheme is therefore temporally
first order. The time level $**$ indicates that we still have to
perform an update for the electron-radiation energy exchange. The backward
Euler equations for the radiation group energies $E_{\rm g}$ can be written as
\begin{equation}
  \frac{E_{\rm g}^{n+1}-E_{\rm g}^*}{\Delta t} = \sigma_{\rm g}^*(B_{\rm g}^* - E_{\rm g}^{n+1})
  + \nabla\cdot D_{\rm g}^*\nabla E_{\rm g}^{n+1},
  \label{eq:splitgroupenergy}
\end{equation}
in which $\sigma_{\rm g}$ is proportional to the group mean Planck opacity,
while $B_{\rm g}$ is the group energy of the blackbody radiation.
The radiation diffusion coefficient $D_{\rm g}$ depends on the group mean
Rosseland opacity. Our multigroup model is flux limited and the flux
limiter is incorporated in $D_{\rm g}$. By introducing the change in the
electron temperature $\Delta T_e=T_e^{**} - T_e^*$ and radiation group
energies $\Delta E_{\rm g} = E_{\rm g}^{n+1}-E_{\rm g}^*$, Eqs.
(\ref{eq:splitionenergy})--(\ref{eq:splitgroupenergy}) can be combined into
equations for these changes
\begin{eqnarray}
  \left[ \frac{C_{Ve}^*}{\Delta t} + \sigma_{ie}'
    - \nabla\cdot C_e^*\nabla \right] \Delta T_e
  &=& \sigma_{ie}'(T_i^* - T_e^*) + \nabla\cdot C_e^*\nabla T_e^*,
  \label{eq:splitelectronnoncos} \\
  \left[ \frac{1}{\Delta t} +\sigma_{\rm g}^* - \nabla\cdot D_{\rm g}^*\nabla \right]
  \Delta E_{\rm g} &=& \sigma_{\rm g}^*(w_{\rm g}^*B^* - E_{\rm g}^*)
  + \nabla\cdot D_{\rm g}^* \nabla E_{\rm g}^*,\label{eq:splitgroupnoncons}
\end{eqnarray}
where we exploited that Eq. (\ref{eq:splitionenergy}) is point-implicit,
which results in the modified ion-electron couplings coefficient
$\sigma_{ie}' = \sigma_{ie}/(1+\Delta t \sigma_{ie}/C_{vi})$.
The $w_{\rm g} = B_{\rm g}/B_e$ is the Planck weight. To obtain
a discretized set of equations, we use the scheme proposed in
\cite{vanderholst2012}. This numerical scheme is overall spatially
second-order, consistent, and conservative. Here it suffices to mention
that once numerical solutions are obtained for these equations by means of a
linear solver, the energy densities for the radiation groups, ions, and
electrons are updated using
\begin{eqnarray}
  E_{\rm g}^{n+1} &=& E_{\rm g}^* + \Delta E_{\rm g}, \\
  E_i^{n+1} &=& E_i^* + \Delta t \sigma_{ie}' (T_e^{**} - T_i^*), \\
  E_e^{n+1} &=& E_e^* + C_{Ve}^*(T_e^{**} - T_e^*)
  + \Delta t \sum_{g=1}^G \sigma_{\rm g}^* (E_{\rm g}^{n+1}-w_{\rm g}^*B^*).
  \label{eq:conseeupdate}
\end{eqnarray}
These updates conserve the total energy to round-off errors.
We note that we changed the Eqs. (\ref{eq:splitionenergy}) and
(\ref{eq:splitelectronenergy}) to solve for the temperatures $T_i$ and
$T_e$ instead of the plackian quantities $B_i=aT_i^4$ and $B_e=aT_e^4$ as
described in Ref. \cite{vanderholst2011}. This improves the scheme at low
temperatures, since we no longer have divisions by temperatures in this system.

The implicit scheme in CRASH uses a preconditioned Krylov solver: GMRES,
BiCGSTAB, or preconditioned conjugate gradient (PCG) as described in Ref.
\cite{vanderholst2011}. Our original preconditioner is based on the
Incomplete Lower Upper (ILU) decomposition with no fill-in. Since CRASH uses
a block-adaptive grid, we find it advantageous to use a Schwartz type
preconditioning on a block-by-block basis \cite{toth2006}. While the block ILU
preconditioner
gives satisfactory results for small to medium problem size, the number
of Krylov iterations can easily be a few 100s to even 1000s on large
problems due to the stiffness of the linear system. In addition, some of the
radiation groups with small energy
content fail to converge due to round-off errors.

As an alternative to the ILU preconditioner, we have implemented an
interface with an algebraic
multi-grid (AMG) preconditioner into CRASH. We use the BoomerAMG solver
from the HYPRE library \cite{falgout2002}. This preconditioner requires
much fewer Krylov iterations than the ILU preconditioner, but the
AMG iterations are much more expensive. The BoomerAMG has also a setup time
at the beginning of the Krylov solve that is rather significant, and does not
scale well to many processors. Despite these issues, we find in our latest
2-D simulations with the new laser package and improved opacities that using
AMG results in a more accurate solutions than the ILU preconditioner. HYPRE
also allows us to set more demanding tolerance criteria for the implicit
solver and still obtain converged solution in a small number of iterations.
For the typical laser-driven radiative shock simulations of the CRASH project
it turns out that BoomerAMG of HYPRE outperforms the ILU preconditioner in
computational time.

\section{Material interface}\label{sec:interface}

\begin{figure}
{\resizebox{0.48\textwidth}{!}{\includegraphics[clip=]{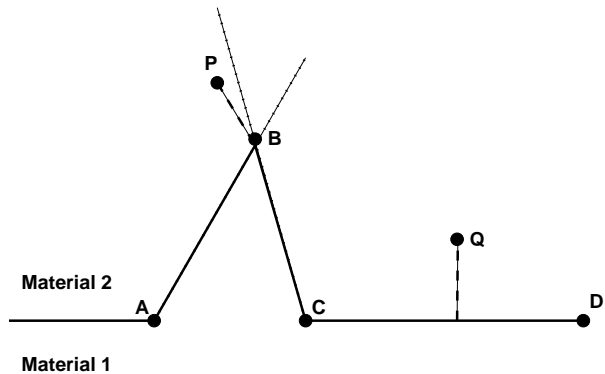}}}
\caption{A material interface defined by the segments AB, BC, and CD. Points
P and Q belong both to material 2.}
\label{fig:interface}
\end{figure}

Figure \ref{fig:interface} shows two different materials separated by an
interface consisting of 4 straight segments. For every point in the plane, we
find the closest segment. For example, for point Q the closest segment is CD.
Since Q is on the left side of the CD vector (directed from point C to D), it
belongs to material 2, and its level set function is set by the normal
distance to segment CD, indicated by the dashed line. For point P, however,
there are two closest segments: AB and BC. P lies on the left side of the AB
vector, which corresponds to material 2,
but on the right side of the BC vector, which would indicate material 1. In
general there can be several interface segments at the same minimal distance
from point P.  One can show that it is always the segment that has the
{\it largest} normal distance from the point, in this case segment AB, that
signals the correct material.

\bibliographystyle{elsarticle-num}

\end{document}